\documentclass[twocolumn,prb,prl,nobibnotes,altaffilletter,amsmath,amssymb,amsfonts]{revtex4}
\date{\today}

\usepackage[latin1]{inputenc}
\usepackage{graphicx}
\usepackage{amsmath}
\usepackage{latexsym}
\usepackage{epsfig}

\begin{document}

\title{Unique Dynamic Correlation Length in Supercooled Liquids}
\author{Claudio Maggi}
\email{cmaggi@ruc.dk}
\author{Bo Jakobsen}
\email{boj@ruc.dk}
\author{Jeppe C. Dyre}
\email{dyre@ruc.dk}
\affiliation{DNRF Centre ``Glass and Time'', IMFUFA, Department of Sciences, Roskilde University,
Postbox 260, DK-4000 Roskilde, Denmark}

\begin{abstract}
We present a direct comparison of the number of dynamically correlated molecules  in the shear-mechanical and dielectric relaxations of the following seven supercooled organic liquids: triphenylethylene, tetramethyl-tetraphenyl-trisiloxane, polyphenyl ether, perhydrosqualene, polybutadiene, decahydroisoquinoline, and tripropylene glycol. For each liquid we observe that the numbers of dynamically  correlated molecules in the shear and in the dielectric relaxation are proportional. 
We show that this proportionality can be explained by the constancy of the decoupling index of the shear and dielectric relaxation times in conjunction with time-temperature superposition. Moreover the value of this proportionality constant is related to the difference in stretching of the shear and dielectric response functions. The most significant deviations from unity of this constant are found in a liquid with strong hydrogen bonds and in a polymer.
\end{abstract}

\maketitle

\section{Introduction}
\label{sec:intro}

The dynamical processes taking place in a supercooled liquid are complex.
This is due to the fact that the motion of the molecules of a liquid close to the
glass transition is intrinsically collective.
When the liquid enters this ultra-viscous regime \cite{JeppeRev}, the rearrangement of a particle
involves the motion of many of its neighbors.
The idea that the dynamics becomes more and more cooperative
has led to searches for a growing length scale as the dynamics slow-down upon cooling. So far no standard static correlation has revealed a detectable growing correlation length arising in the supercooled phase.
On the other hand, \emph{dynamic correlations} \cite{Ediger} may account for the evolution of the correlation length scales involved in the glass transition. Recently Berthier, Biroli and co-workers invented a simple and powerful method to estimate the four-point susceptibility, $\chi_4$ \cite{BeScie, BeJChem1, BeJChem2}. The central idea is to estimate the four-point function via a more accessible three-point function. The function $\chi_4$, which cannot be easily measured directly, can be approximated from the temperature evolution of any
 measured dynamic variable.

We can measure the frequency-dependent dielectric susceptibility, $\epsilon(\omega)$, and shear modulus, $G(\omega)$ using the same cryostat and covering overlapping temperature-frequency ranges \cite{Brian, Brian2, RevSci}. The \emph{piezo-shear-gauge} (PSG) technique \cite{RevSci} allows us to measure the dynamic shear modulus of a supercooled liquid close to its glass transition (where $G$ typically assumes values between 0.1 MPa and 10 GPa) in a wide frequency range ($10^{-3}-10^{4}$ Hz). In this work we extract and compare the number of dynamically correlated molecules in the structural (alpha) relaxation from two different dynamic variables: the dielectric susceptibility and the shear modulus. 

\begin{table}
\begin{tabular}{|l|c|c|c|c|}
\hline
 & $Tg$ [K] & $m$ & $I(T)$ & (clear) $\beta$-relaxation\\
\hline
TPE & 249 & 73 & 3.4 -- 3.5 & no \\
\hline
DC704 & 211 & 83 & 3.7 -- 3.9 & no\\
\hline
PPE & 245 & 80 & 3.9-3.9 & no\\
\hline
Squalane & 167 & 64 & 0.4 -- 2.9 & yes\\
\hline
PB20 & 176 & 79 & 3.7 & yes\\
\hline
DHIQ & 179 & 154 & 3.8 -- 8.3 & yes\\
\hline
TPG & 190 & 65 & 1.4 -- 3.0 & yes \\
\hline
\end{tabular}
\caption{Properties of the seven liquids studied (from Refs. \cite{Kr&Bo, Bothesis, DiMarzio, Tina, Repos}). $T_g$ is the glass transition temperature, $m$ is the Angell fragility, $I$ is the temperature index (the values of $I$ are reported for the highest and lowest temperature studied). The last column indicates if the liquid has or not a clear secondary $\beta$-relaxation. All the data here refer to dielectric measurements.
\label{tab:prop}}
\end{table}

We analyze below dielectric and shear-dynamic data collected and published by our group \cite{Kr&Bo, DiMarzio}, available on-line \cite{Repos}.
This study focuses on seven liquids: triphenylethylene (``TPE''), tetramethyl-tetraphenyl-trisiloxane (``DC704''), polyphenyl ether (``PPE''), perhydrosqualene (``squalane''), polybutadiene (``PB20''),
decahydroisoquinoline (``DHIQ''), and tripropylene glycol (``TPG''). DC704, TPE, PPE, squalane and DHIQ are molecular van der Waals bonded liquids, TPG has hydrogen bonds, and PB20 is a polymer with molecular weight of 5000 g/mol. All liquids were used as acquired. The PPE used is the Santovac\textregistered  5 vacuum pump fluid, and DC704 is the Dow Corning\textregistered 704 diffusion pump fluid. All the other liquids were acquired from Sigma-Aldrich. All the experimental details about these measurements can be found in Refs.\ \cite{Kr&Bo,Brian, Brian2, RevSci}.

Some properties of the liquids \cite{Kr&Bo, Bothesis, DiMarzio, Tina, Repos} are reported in Table \ref{tab:prop}. Here the relaxation time $\tau$ is defined by the inverse loss peak frequency and the glass transition temperature $T_g$ is defined as the temperature where the loss-peak is located at $2 \pi 10^{-3}$ rad/s. The temperature dependence of the relaxation time around $T_g$ is expressed in terms of the Angell fragility index \cite{fragility1, fragility2, fragility3, fragility4} 

\begin{equation}\label{eq:frag}
m=\left. \frac{d \log_{10}\tau}{d (T_g/T)} \right|_{T=T_g} 
\end{equation}

\noindent The temperature dependence of activation energy is quantified via the temperature index \cite{index, Tina} 

\begin{equation}\label{eq:index}
I(T)=\frac{d \ln \Delta E(T)}{d \ln T}
\end{equation}

\noindent where $\Delta E(T)$ is the activation energy defined by $\tau=\tau_0\exp(\Delta E(T)/k_B T)$. Table \ref{tab:prop} reports the variation of $I(T)$  in the temperature interval studied \cite{Tina}. Table \ref{tab:prop} also reports presence of a clear Johari-Goldstain $\beta$-relaxation \cite{beta} in the dielectric spectrum of the liquid.

\section{Comparison of shear-mechanical and dielectric responses}
\label{sec:comR}

The studies carried out by our group \cite{Kr&Bo, DiMarzio,
ShePa, Mono} focused on the
temperature-dependence of the the shear-mechanical and dielectric
$\alpha$ relaxation times (indicated with $\tau_G$ and
$\tau_\epsilon$, respectively). Furthermore, comparison of the shape of these two relaxation functions was
presented. The main conclusions of these studies may be summarized as follows:

\noindent {}

(\emph{i}) The relaxation time of the shear modulus  is
generally different from that of the dielectric
susceptibility at the same temperature $T$. The shear-mechanical 
relaxation is always slightly faster than the dielectric, $\tau_\epsilon (T) \geq \tau_G (T)$. 
Nevertheless, the shear and dielectric characteristic alpha relaxation times
evolve in a rather similar way in the liquids studied when $T$ is
changed. This was discussed in detail in \cite{Kr&Bo} where the
\emph{decoupling index} $\tau_\epsilon (T)/\tau_G (T)$
was reported and its insignificant temperature dependence was established
($\tau_\epsilon (T)/\tau_G (T) \simeq
\mathrm{const}$). This picture is also confirmed by other studies 
found in the literature \cite{sd1, sd12, sd2, sd3, sd4}.

\noindent {}

(\emph{ii}) The shear response function and the dielectric response function
generally have different shapes. In liquids that do not show
any detectable Johari-Goldstain $\beta$-relaxation \cite{beta} the
shape of each frequency-dependent response is found to be almost
temperature independent. This feature is referred as
\emph{time-temperature superposition} (TTS), and it is found to  hold
to a very good degree in the temperature-frequency range
explored \cite{Kr&Bo, ShePa} both for the shear and the dielectric
relaxation \cite{Albena}.

\noindent {}

(\emph{iii}) For those liquids that have a clear beta-relaxation the 
alpha relaxations in the shear and
dielectric spectrum seem to approach a temperature independent shape as
the temperature is lowered. This has been presented in detail in Refs.
\cite{Kr&Bo, ShePa, Mono, Albena} suggesting
that for the alpha process alone TTS applies, while in the full
spectrum TTS is lost because of the presence of the beta process.

\noindent {}

In the following we show that (\emph{i}) and (\emph{ii}) imply that the shear and the dielectric numbers of dynamically correlated molecules (for the liquids without clear beta relaxation) are proportional in the temperature range studied. Moreover, the same conclusion applies if we \emph{assume} TTS (as suggested by (\emph{iii}) and also done in \cite{Simone})
to hold for the alpha process in those liquids that have a secondary
relaxation. To understand this link it we first briefly recall
how to approximate the four-point susceptibility.

\section{ Estimation of The Number of Dynamically Correlated Molecules}
\label{sec:calX}

The four-point correlator can be interpreted as the variance of the dynamics around its average value. One can estimate this function from the following equation
(Refs. \cite{BeScie, BeJChem1, BeJChem2, Ladi, Cecile, Simone})

\begin{equation} \label{eq:chi4}
\chi_4(\omega,T) \simeq \frac{k_B}{c_P} \left( \frac{\partial\tilde{\chi}'(\omega,T)}{\partial \ln T} \right)^2.
\end{equation}

\noindent In this equation $\tilde{\chi}'$ is the normalized real
part of the response function and $c_P$ is the
configurational heat capacity per molecule at constant pressure. The
right-hand side of Eq. (\ref{eq:chi4}) is an approximation of
$\chi_4$, it actually represents a lower bound for this function.
Nevertheless, this method is found to give values of the four-point susceptibility 
in good agreement with the actual values of $\chi_4$
when these can be evaluated directly (for example in computer
simulations) \cite{BeJChem1, BeJChem2}. The characteristic value of
the four-point function (i.e., the typical number of correlated molecules
in the relaxation, $N_\text{corr}$) is associated with the maximum of
$\chi_4$,

\begin{equation} \label{eq:N}
N_{\text{corr}}(T)=\max_{\omega}[\chi_4(\omega,T)].
\end{equation}

\noindent The maximum of this function is consistently found
to close to a frequency close to the loss-peak frequency of the alpha dynamics.

The normalized response function $\tilde{\chi}'$ appearing in Eq.
(\ref{eq:chi4}) is computed by subtracting a baseline parameter to the
measured response (for example the dielectric susceptibility), 
subsequently dividing by the amplitude of the function \cite{Ladi}

\begin{equation} \label{eq:norm}
\tilde{\chi}(\omega,T)=\frac{\chi(\omega,T)-\chi_\infty}{\Delta \chi}.
\end{equation}

In this work we fit all response functions with the
Havriliak-Negami (HN) function \cite{HN}

\begin{equation} \label{eq:HN}
\chi(\omega,T)=\chi_{\infty}+\frac{\Delta \chi}{[1+(i\tau \omega)^\alpha]^\beta}.
\end{equation}

\begin{figure}
\begin{center}
\includegraphics[width=9.5cm]{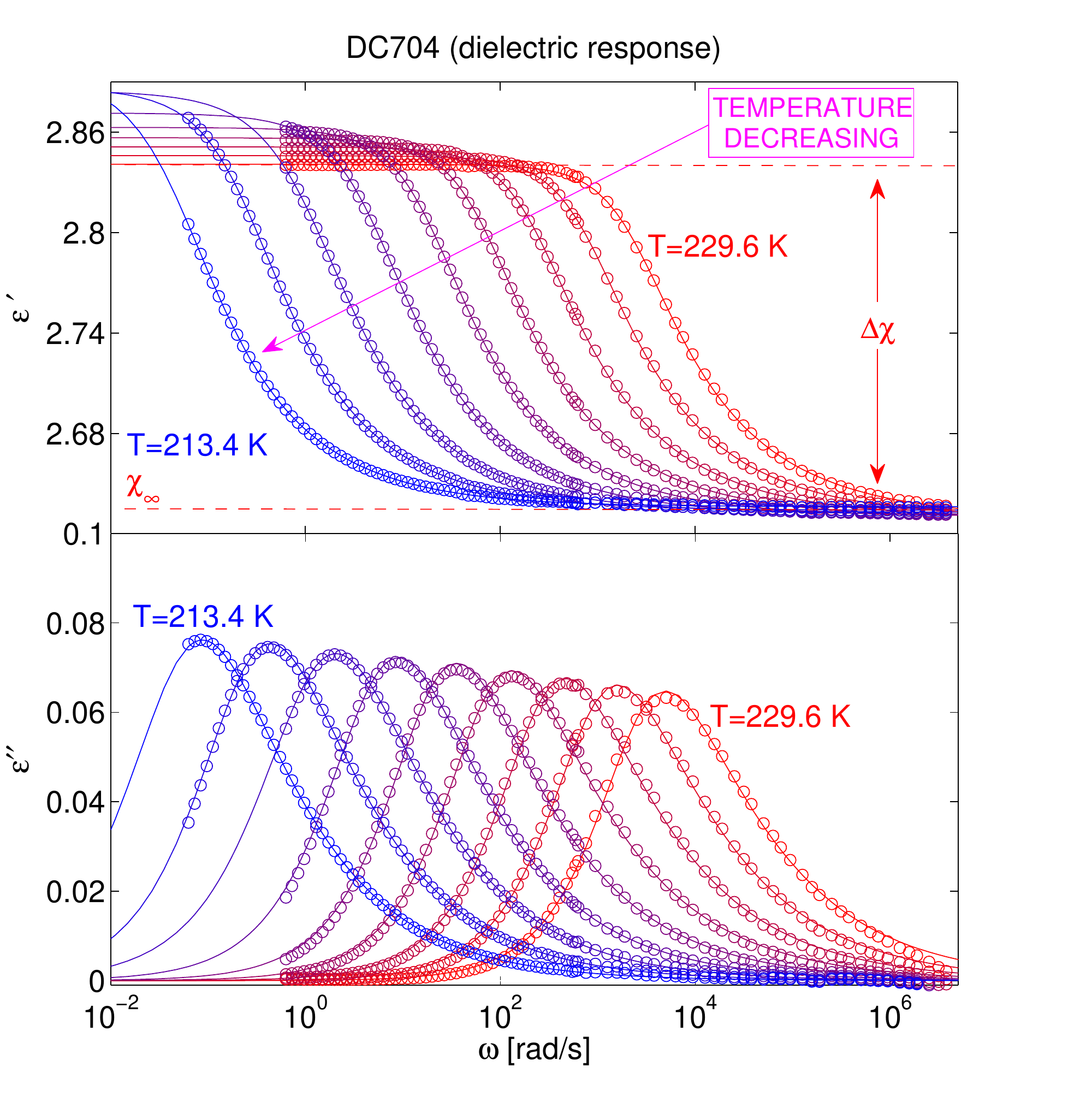}
\end{center}
\caption{Real and imaginary part (top and bottom respectively) of
the dielectric response of DC704 measured at $T=$229.6,  227.6
225.5,  223.5,  221.5,  219.5,  217.5,  215.4,  213.4 K (open
circles). The full lines are fits with
the HN function Eq. (\ref{eq:HN}). The amplitude $\Delta\chi$ and
the baseline $\chi_\infty$ of the fitting function are shown
graphically for the curve at $T=$229.6 K.} \label{fig:FITS}
\end{figure}

\begin{figure*}
\begin{center}
\includegraphics[width=15cm]{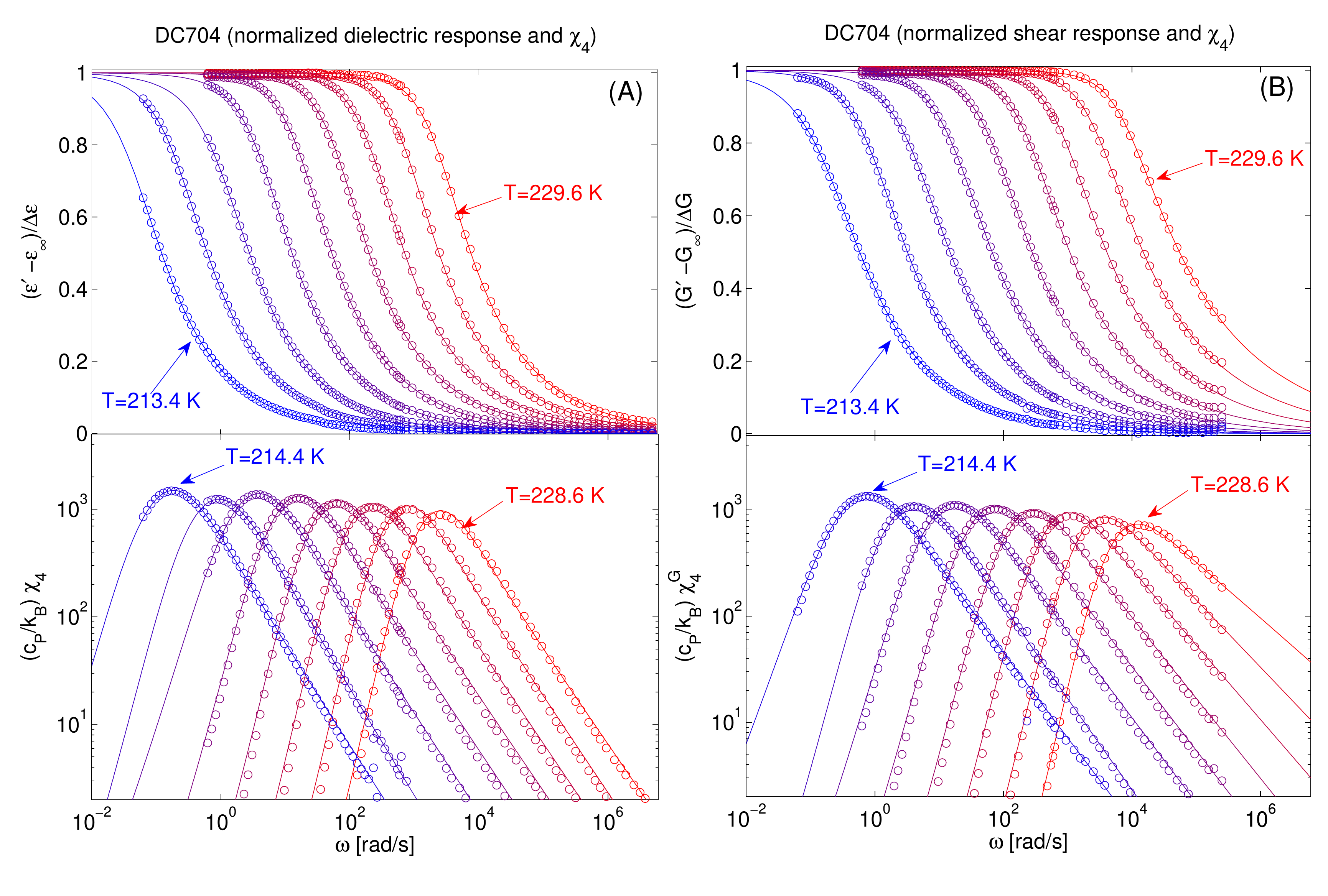}
\end{center}
\caption{(\textbf{A-Top}) Normalized real part of the dielectric response for
DC704 at $T=$ 229.6,  227.6  225.5,  223.5,  221.5,  219.5,
217.5,  215.4 and 213.4 K. The full lines are the corresponding
(normalized) fits to the HN functions. (\textbf{A-Bottom}) $(c_P/k_B) \chi_4=\left({\partial\tilde{\chi}'(\omega,T)}/{\partial \ln T} \right)^2$ for
the dielectric (left) and shear (right) relaxation at T= 228.6,
226.6,  224.5,  222.5,  220.5, 218.5, 216.4 and  214.4 K. The
full line are the corresponding $(c_P/k_B) \chi_4$ computed form the
fitting HN functions. (\textbf{B}) Same as A for the shear-mechanical response} 
\label{fig:allNCHI}
\end{figure*}

\noindent The Appendix details how we introduce the
assumption of TTS in the analysis of the spectra of the liquids
presenting a beta process (this is done by fixing the $\alpha$ and $\beta$ parameters of the function (\ref{eq:HN})). An example of the fitting is reported in Fig. \ref{fig:FITS} for the dielectric responses of DC704. The normalized functions are shown in the upper parts of Fig. \ref{fig:allNCHI}.A (dielectric) and in Fig. \ref{fig:allNCHI}.B (shear). 

\noindent The function $\chi_4$ obtained from Eq. \ref{eq:chi4}
is shown in Fig. \ref{fig:allNCHI} for the dielectric responses 
and the shear-mechanical response of DC704.
Note that the maximum of these functions at the same temperature
is located at different frequencies in the shear and the dielectric case (as is also the case for the loss peaks of the responses). Moreover, the shape of $\chi_4$ is slightly different in the shear and in the dielectric case as discussed in detail below.

Once we have determined the maximum of $\chi_4$, the quantity $N_\text{corr}$
can be obtained via Eq. (\ref{eq:N}).
In this way two independent estimates of the number of dynamically
correlated molecules can be obtained: the number
of correlated molecules in the shear relaxation $N_G$ and in the
 dielectric relaxation $N_\epsilon$. Note that for comparing these
  two numbers knowledge of $c_P$ is unnecessary, being only a constant
  multiplicative factor in Eq. (\ref{eq:N}).

\begin{figure}
\begin{center}
\includegraphics[width=9.75cm]{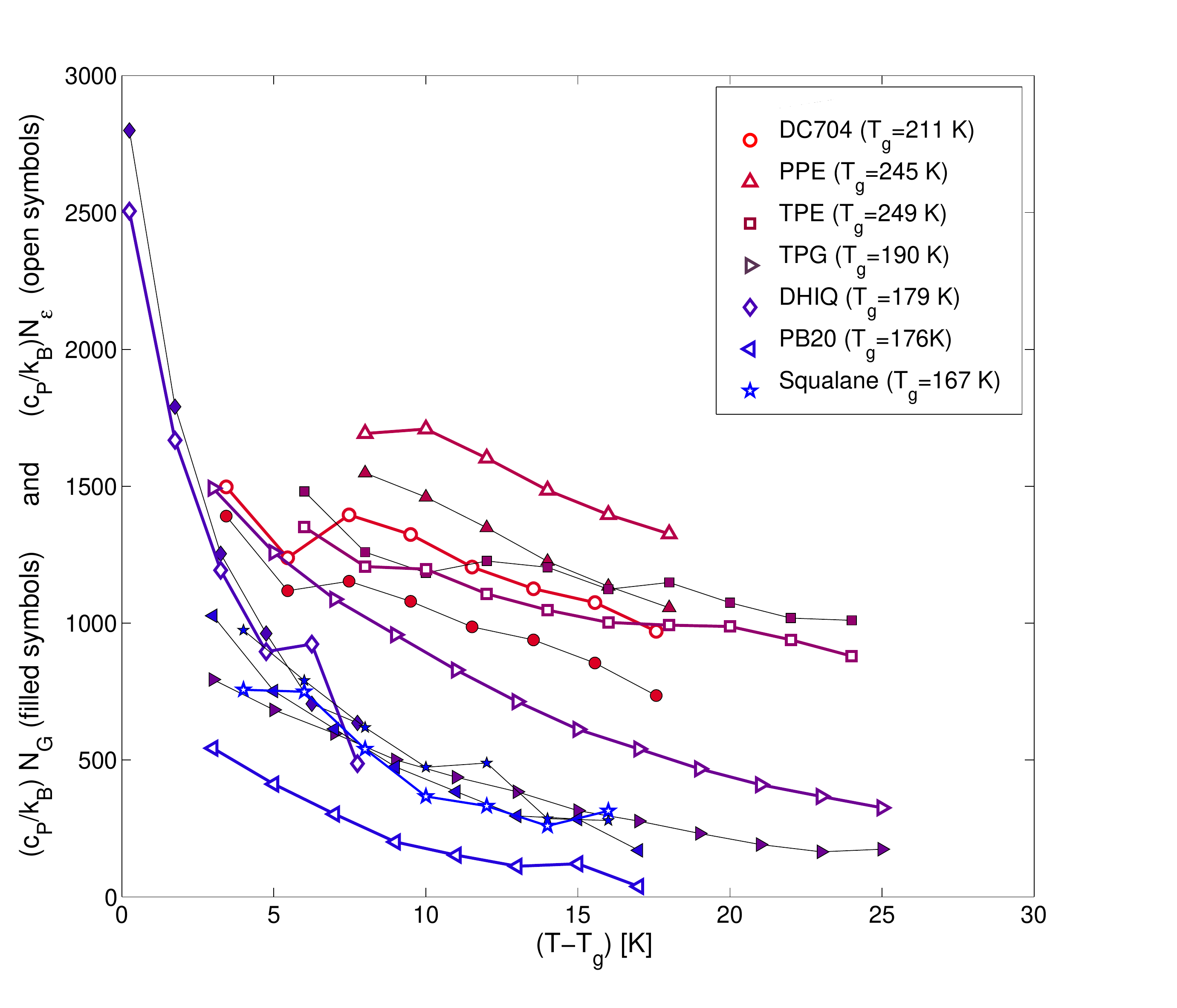}
\end{center}
\caption{The quantity $(c_P/k_B) N_\text{corr}$ for shear (filled symbols) and dielectric (open symbols) as a function of $(T-T_g)$ for the liquids studied (see legend). $T_g$ is the glass transition temperature for the dielectric relaxation from Ref.\ \cite{Kr&Bo} (see legend and Table \ref{tab:prop}).}
\label{fig:NvsT}
\end{figure}

From Fig. \ref{fig:NvsT} we can appreciate the growth of the shear and diellectric $N_\mathrm{corr}$ upon cooling.  The minimum increase of $N_\epsilon$ is of a factor
$\sim$1.6 in TPE and its maximum increase is of a factor $\sim$5.7
found in TPG. The relaxation times of the responses
studied grow at least four orders of magnitude in all
liquids.

Let us now see the form assumed by the equations (\ref{eq:chi4}) and (\ref{eq:N}) if TTS applies. To do this let us consider the very general expression for a (normalized) response function obeying TTS:

\begin{equation} \label{eq:CC}
\tilde{\chi}(\omega,T)=\phi(\omega \tau(T)).
\end{equation}

\noindent This is the case of Eq. \ref{eq:HN} if the 
parameters $\alpha$ and $\beta$ are kept constant. Differentiating
the real part of Eq. (\ref{eq:CC}) with respect to $\ln T$ (as in Eq. (\ref{eq:chi4})) we obtain

\begin{equation} \label{eq:CCn}
\frac{\partial\phi'(\omega \tau(T))}{\partial \ln T}=
\omega \frac{d \phi'(x)}{d x} \frac{\partial \tau(T)}{\partial \ln T}.
\end{equation}

\noindent where the prime indicates the real part and we have introduced $x\equiv\omega \tau(T)$. The maximum of the function (\ref{eq:CCn})  can be estimated setting $\omega=\tau^{-1}$ (a minor correction term is present in the case of very large stretching \cite{Simone}):

\begin{equation} \label{eq:CCnmax}
N_\text{corr} \propto \left[\max_\omega \left({\frac{\partial\phi'(\omega \tau(T))}{\ln T}}\right) \right]^2=
f^2(1) \left( \frac{\partial \ln \tau(T)}{\partial \ln T} \right)^2
\end{equation}

\noindent where $f(x)=(d \phi' (x) /d x)$. From Eq. (\ref{eq:CCnmax}) it is clear that the growth of $N$ is determined uniquely by the growth of the relaxation time upon cooling if TTS strictly holds. If (as stated in (\emph{i})) the decoupling index has a negligible temperature-dependence
($\tau_\epsilon(T)/\tau_G(T) \simeq \mathrm{const}$) then

\begin{equation}
\label{eq:dtau}
{\left( \frac{\partial \ln \tau_\epsilon (T)}{\partial \ln T} \right)}^2 \simeq
{\left( \frac{\partial \ln \tau_G (T)}{\partial \ln T} \right)}^2.
\end{equation}

\noindent This means that the decoupling index of the characteristic number
of correlated molecules in the shear and dielectric relaxation is also constant as $T$ is lowered:

\begin{equation}
\label{eq:dcN}
\frac{N_\epsilon(T)}{N_G(T)} \simeq \mathrm{const}
\end{equation}

\noindent where the constant is determined by the stretching of the shear and
dielectric relaxations. In other words, the growth of $N_G$ and $N_\epsilon$
is identical upon cooling, while their difference in absolute values is set by the different
 (temperature independent) shape of the two response functions.

\section{Comparison of The Shear and Dielectric Numbers of Dynamically Correlated Molecules}

\begin{figure}
\begin{center}
\includegraphics[width=9.75cm]{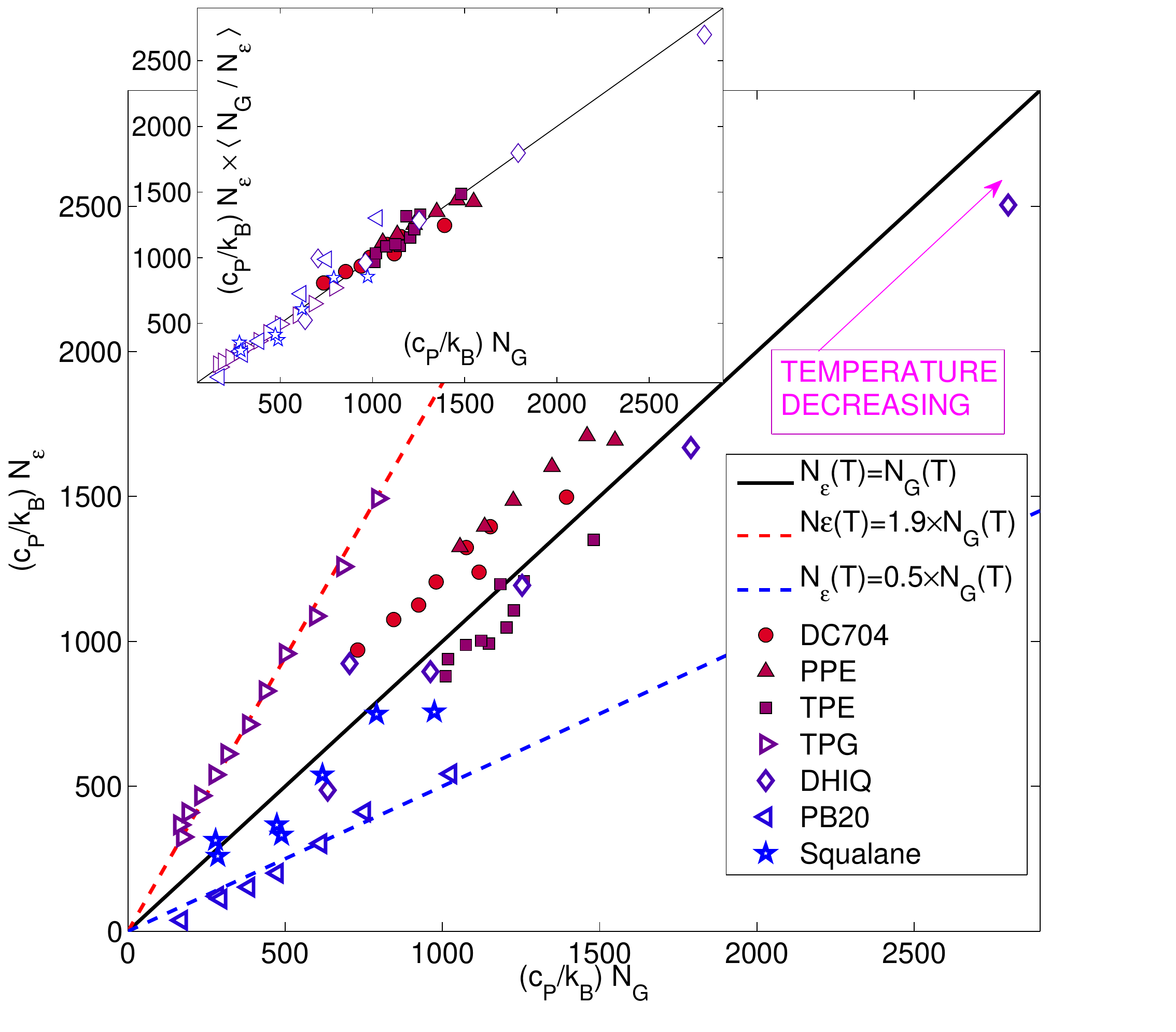}
\end{center}
\caption{Characteristic number of correlated molecules in the dielectric relaxation
plotted versus the characteristic number of correlated molecules in the shear mechanical relaxation
at the same temperatures for the seven glass-forming liquids (see legend). The full line
indicates $N_\epsilon=N_G$ while the dashed lines represents $N_\epsilon=\lambda N_G$
(with $\lambda=$1.9 and 0.5 respectively for the upper and the lower line). In the inset we show the data collapse on the line $N_\epsilon \propto N_G$ obtained when we multiply $N_\epsilon$ by the value $\langle N_G/N_\epsilon \rangle$ where the average is taken over the temperature range studied.}
\label{fig:NEvsNG}
\end{figure}

The results expected from the constant decoupling index and TTS can be readly checked. 
Plotting $N_\epsilon(T)$ versus $N_G(T)$ as in Fig. \ref{fig:NEvsNG} we check that these quantities approximatively differ only by a multiplicative factor in
 the temperature range studied for all liquids considered in this work. As seen in the inset of Fig. \ref{fig:NEvsNG} all the data collapse onto the line $N_\epsilon \propto N_G$ if we multiply $N_\epsilon$ by the value $\langle N_G/N_\epsilon \rangle$ where the average is taken over the temperature range studied.

\begin{figure}
\begin{center}
\includegraphics[width=9.75cm]{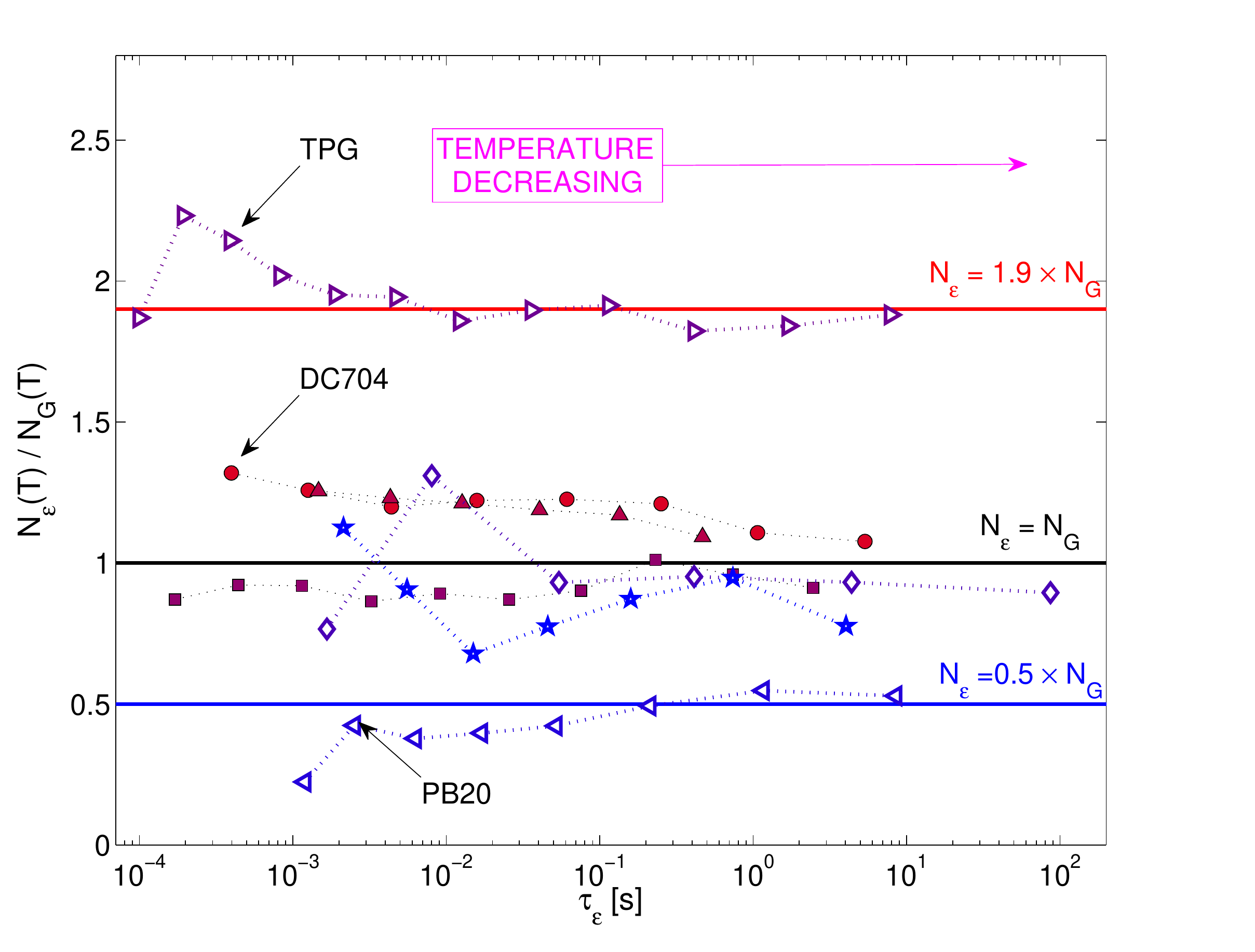}
\end{center}
\caption{Ratio between the characteristic number of dynamically correlated molecules
in the shear and in the dielectric relaxation at the same temperature as
a function of the (dielectric) relaxation time for different liquids (same symbols as
 in Fig. \ref{fig:NEvsNG}). The full lines represent
 $N_\epsilon=\lambda N_G$ with $\lambda=1.9$,1,0.5 respectively from top to bottom.}
\label{fig:ratios}
\end{figure}

 We test Eq. (\ref{eq:dcN}) directly in Fig. \ref{fig:ratios}. As seen from this figure the
  decoupling index $N_\epsilon(T)/N_G(T)$ of the number of correlated molecules in the dielectric
  and shear relaxation is very weakly temperature dependent and it does not
  show any clear trend of a systematic increase or decrease. Note that, while $N_\epsilon(T)/N_G(T)$ stays
constant, $N_\epsilon$ and $N_G$ both grow significantly upon cooling
for all liquids (see Fig. \ref{fig:NvsT}).

\begin{figure}
\begin{center}
\includegraphics[width=9.75cm]{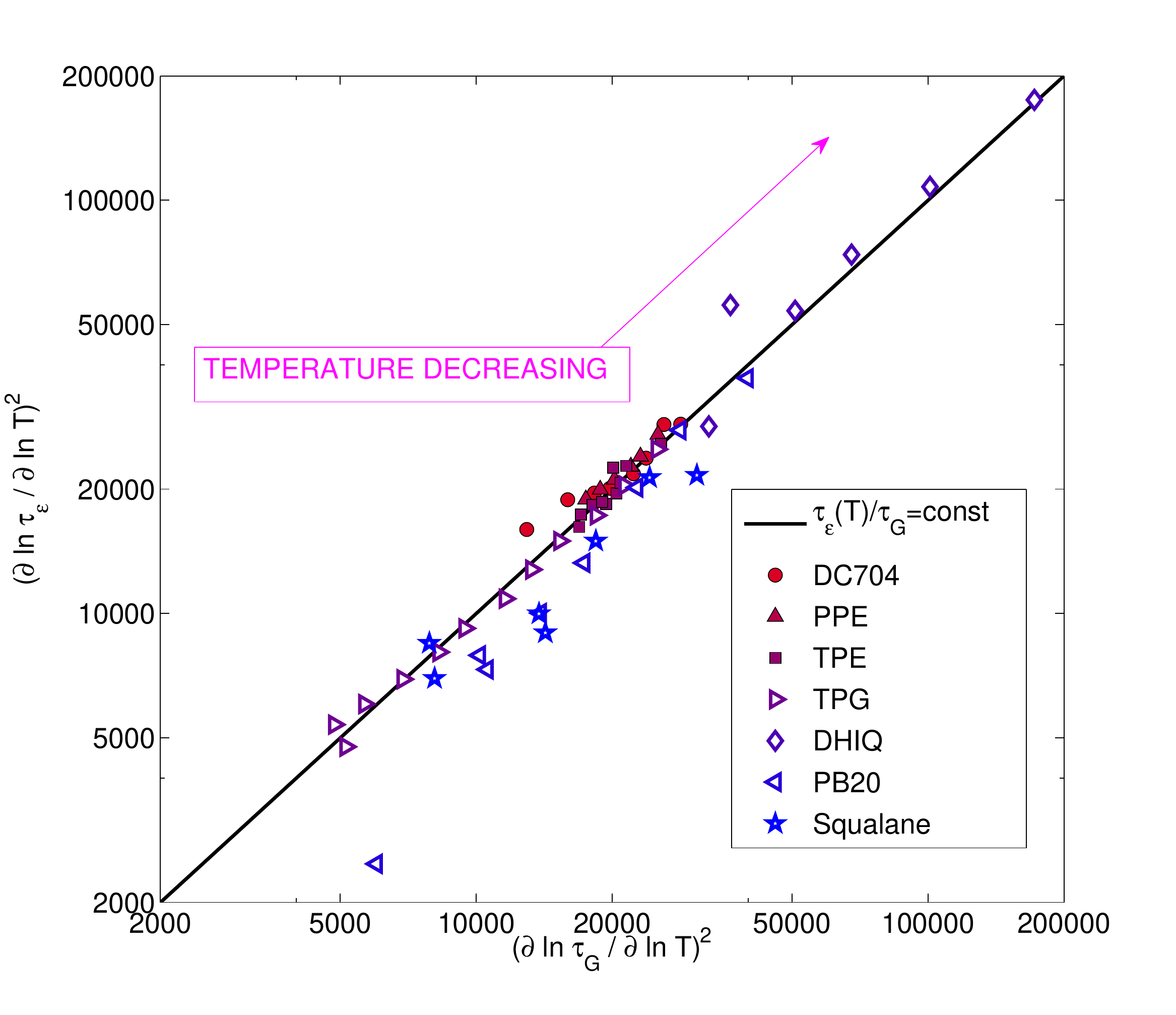}
\end{center}
\caption{$ (\partial \ln \tau_\epsilon/\partial \ln T )^2$ plotted versus
 $ (\partial \ln \tau_G/\partial \ln T )^2$ for the seven liquids (see legend).
 The full line represents $ (\partial \ln \tau_\epsilon/\partial \ln T )^2=
 (\partial \ln \tau_G/\partial \ln T )^2$ (Eq. \ref{eq:dtau}).}
\label{fig:dtau}
\end{figure}

In Fig. \ref{fig:dtau} we test further the constancy of the decoupling index.
Fig. \ref{fig:dtau} demonstrate the validity of Eq. (\ref{eq:dtau}). The equality $(\partial \ln \tau_\epsilon/\partial \ln T)^2 = (\partial \ln \tau_G/\partial \ln T )^2$ seems to hold to a good approximation as expected from 
$\tau_\epsilon (T)/\tau_G (T) \simeq
\mathrm{const}$.

We stress once again that the multiplicative factor between the shear and dielctric $N_\text{corr}$ (i.e., the constant appearing in Eq. (\ref{eq:dcN})) is determined by the shape of the shear and dielectric response. Indeed, the function $f(1)$ defined in Eq. (\ref{eq:CCnmax}) depends on the form of the relaxation functions that set $N_\epsilon/N_G\simeq[f_\epsilon(1)/f_G(1)]^2$. When the response is modelled by the HN function (\ref{eq:HN}), the value of $f(1)$ depends only on the parameters $\alpha$ and $\beta$:

\begin{equation}
\label{eq:fone}
f(1)= -\alpha\beta \, \text{Re} [i^\alpha/(1+i^\alpha)^{1+\beta}].
\end{equation}

\begin{figure*}
\begin{center}
\includegraphics[width=14.5cm]{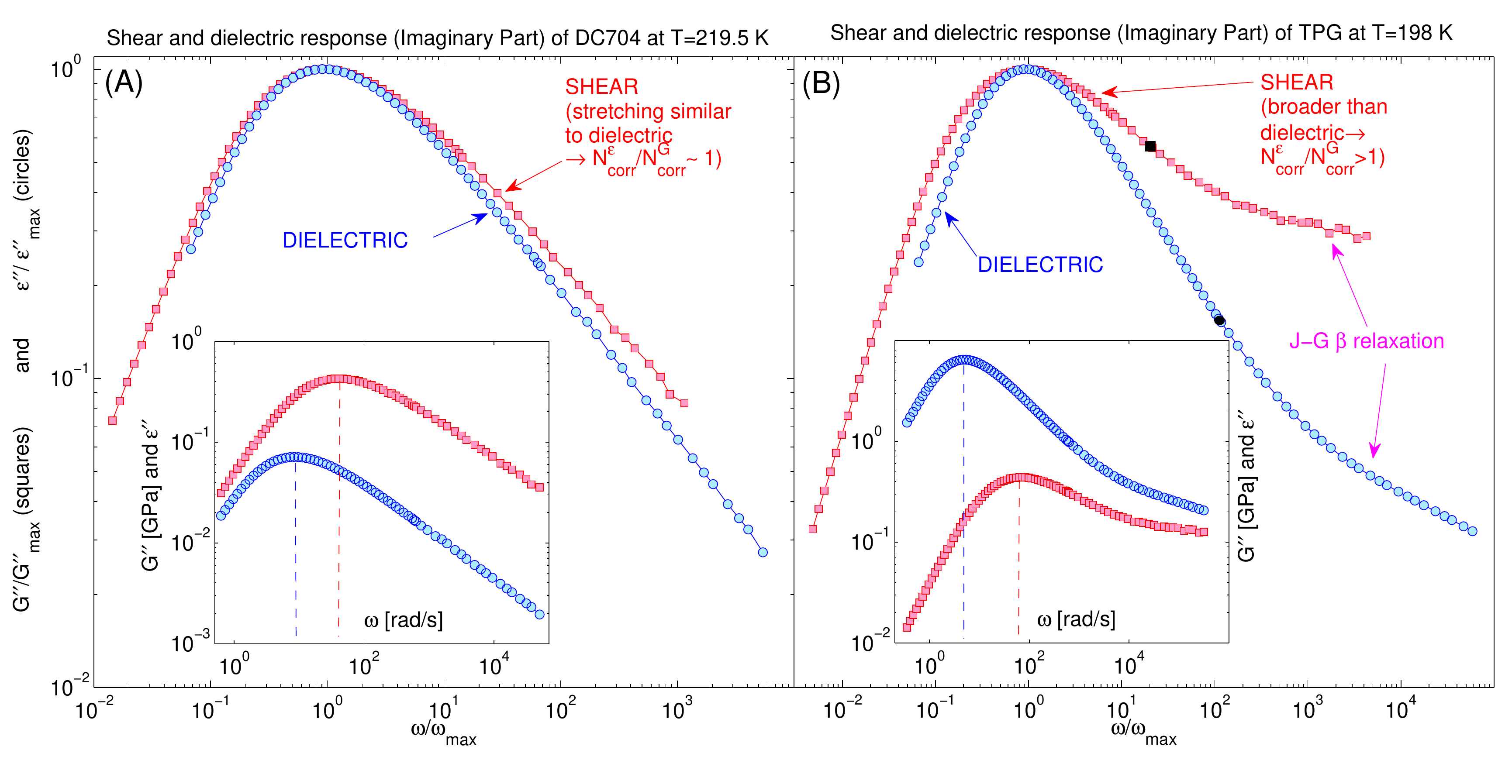}
\end{center}
\caption{(\textbf{A}) Comparison of (normalized) imaginary part of the shear and dielectric response function of DC704 at the same temperature $T=$219.5 K (main panel). The amplitude of the functions is normalized to the loss maxima ($\epsilon''_{\max}$ and $G''_{\max}$) and they are plotted vs $\omega/\omega_{\max}$ for an easier comparison of the shapes. In the inset we show the two susceptibilities as they are are measured. (\textbf{B}) Same as (A) for TPG at $T=$198 K. The black symbols indicate the last point included in the fitting (see Appendix).}
\label{fig:confALL}
\end{figure*}

\noindent From  Eq. (\ref{eq:fone}) it is easy to understand that if the dielctric response function has approximatively the same shape as the shear-mechanical one, the constant of Eq. (\ref{eq:dcN}) is close to unity. This is the case of DC704 as can be seen from Figs. \ref{fig:ratios} and \ref{fig:confALL}.A for which Eq. (\ref{eq:fone}) gives consistently $N_\epsilon/N_G\simeq[f_\epsilon(1)/f_G(1)]^2\simeq1.35$ using the values $\alpha$ and $\beta$ obtained from the fitting.

If the dynamic shear modulus is instead much broader than the dielectric response (as shown in Fig. \ref{fig:confALL}.B for TPG), $N_\epsilon/N_G$ is significantly larger than unity (as seen in Fig. \ref{fig:ratios}). Also for this liquid we can check that the value of this ratio is consistent with the equations given above obtaining $N_\epsilon/N_G\simeq1.97$.

\begin{figure*}
\begin{center}
\includegraphics[width=14.5cm]{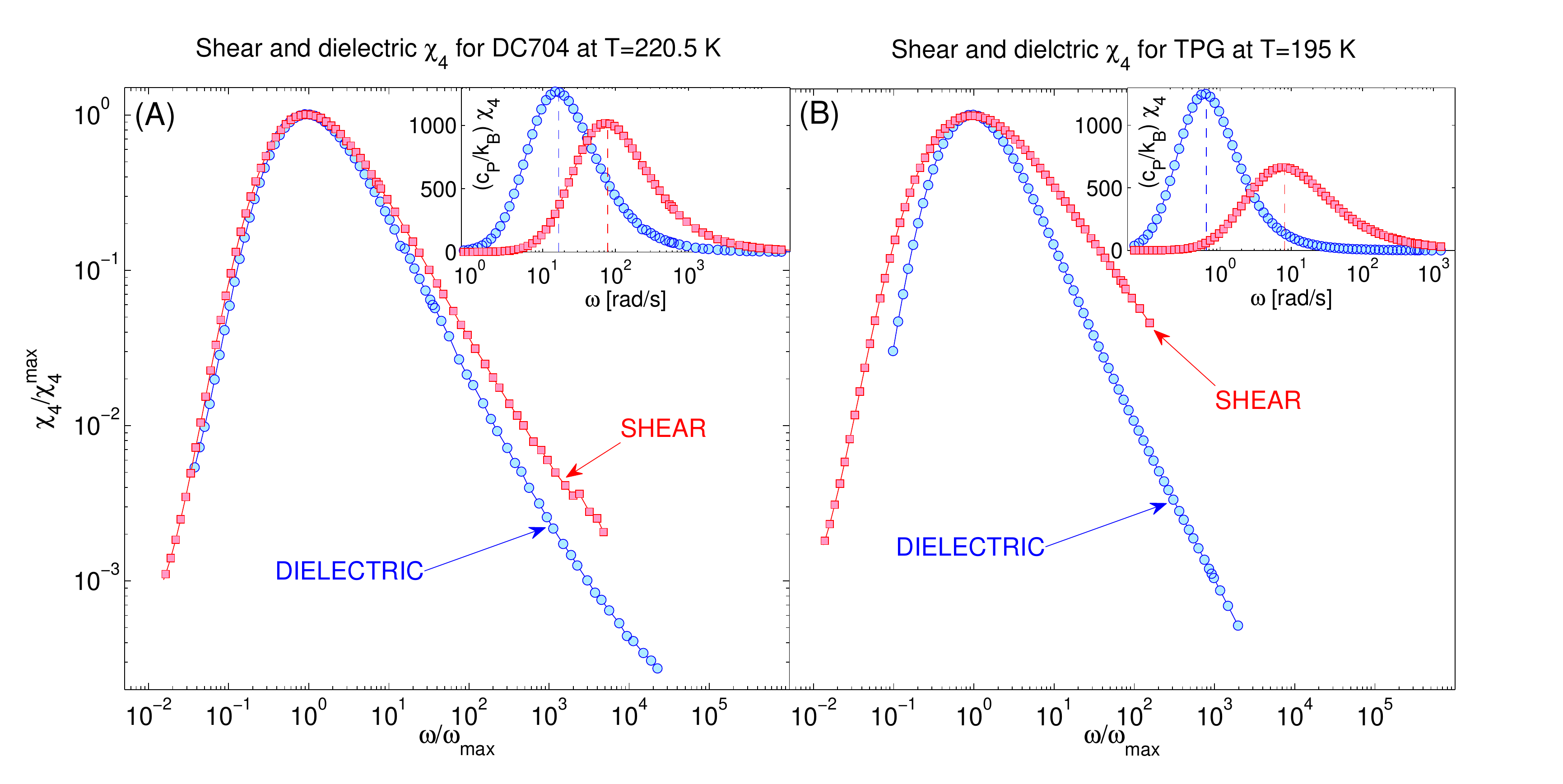}
\end{center}
\caption{(\textbf{A}) Comparison of the shear and dielectric four-point susceptibility of DC704 at the same temperature $T=$220.5 K (main panel). The amplitude of each function is normalized to the maximum ($\chi_4^{\max}$) and it is plotted vs $\omega/\omega_{\max}$ for an easier comparison of the shapes. In the inset we show $\chi_4^G$ and $\chi_4^\epsilon$ calculated from Eq. \ref{eq:chi4}. (\textbf{B}) same as (A) for TPG at $T=$195 K.}
\label{fig:chiALL}
\end{figure*}

Finally we want to stress that the stretching of the relaxation function does not only significantly affect $N_\text{corr}$, but also the full shape of the function $\chi_4(\omega)$ as calculated from Eq. (\ref{eq:chi4}). This is illustrated in Fig. \ref{fig:chiALL}. If the stretching of the shear and dielectric response functions is similar as in DC704 (see Fig. \ref{fig:confALL}.A), $N_\epsilon/N_G$ is close to one, but also the shapes of $\chi_4^G(\omega)$ and $\chi_4^\epsilon(\omega)$ are quite similar as seen in Fig. \ref{fig:chiALL}.A.

If the two response functions have significantly different stretching, the corrresponding four-point susceptibilities will have quite different shapes. This is the case of TPG whose dielctric response function is more ``Debye-like'' than the shear-mechanical one (Fig. \ref{fig:confALL}.B). In this case the functions $\chi_4^G$ is clearly broader than $\chi_4^\epsilon$ as seen in Fig. \ref{fig:chiALL}.B.

\section{Conclusions}

We have compared the shear-mechanical and dielectric characteristic number of dynamically correlated molecules for seven supercooled liquids close to the glass transition.
The number of dynamically correlated molecules
 in the shear-mechanical relaxation is generally different from that of the dielectric relaxation.
Nevertheless, these quantities are approximatively proportional in the explored temperature range. For five of the seven liquids studied the ratio between the shear and dielectric characteristic number of correlated moleculues is close to the unity. The most significant deviations from this unitary ratio are found in a liquid with strong hydrogen bonds and in a polymer. Finally, we showed that the difference in these absolut numbers arises from the different stretching of the dielectric and shear response functions.

Center for viscous liquid dynamics ``Glass and Time'' is sponsored by The
Danish National Research Foundation (DNRF).

\section*{APPENDIX: Analysis details}

Note that, although we consider the real part of $\tilde{\chi}$ when computing the four-point susceptibility, we fit simultaneously
the real and imaginary part with Eq. (\ref{eq:HN}). This is done minimizing the
 (generalized) residual $\chi^2$ for a complex variable $x$, i.e.:
 $ \chi^2=\sum_j(x^j_{exp}-x^j_{th})^\ast \cdot (x^j_{exp}-x^j_{th})$  where the star indicates the complex conjugate.

A further remark on the computation of $\chi_4$ is that Eq. (\ref{eq:chi4})
involves the derivative with respect to the temperature that is, in practice,
performed as finite difference. After obtaining the normalized curves we consider
two successive functions measured at different temperatures $T_1$ and $T_2 (<T_1)$.
Each frequency scan of a response function is carried in a way that each curve has
points in the same frequencies. The derivative appearing in Eq. (\ref{eq:chi4}) is
then computed, at the single frequency, as follows:

$$
\left[ \frac{\partial\tilde{\chi}'(\omega,T)}{\partial \ln T} \right]_{T=(T_1+T_2)/2} \simeq
\frac{\tilde{\chi}'(\omega,T_1)-\tilde{\chi}'(\omega,T_2)}{\ln T_1 - \ln T_2}
$$

Here we illustrate how we introduce the assumption of TTS in the
analysis of the spectra of the liquids presenting a secondary
relaxation process (as TPG, DHIQ, Squalane and PB20). This is done
by fixing the stretching parameters in (\ref{eq:HN}) for the fitting
in the following way.

First we fit the spectrum at the lowest temperatures (with free
stretching parameters) where the secondary process is well separated
from the $\alpha$ relaxation. We exclude from this fit some of the
high frequency data (affected by the secondary relaxation). To
select which data to remove from the minimization we plot the
logarithmic derivative of the imaginary part of the response (that
is $\alpha=\partial \ln \chi'' / \partial \ln \omega$) that shows a
minimum $\alpha_{min}$ at the frequency $\omega_{min}$ where the
$\alpha$ process meets the secondary relaxation. The frequencies
larger than this $\omega_{min}$ are not considered in the fitting.

Once the parameters are found from this low-temperature spectrum
they are fixed to fit all the other spectra up to high temperature.
In those fits we also exclude the high frequency points from the
computation in the same
 way illustrated above.

When this procedure is completed $\chi_4$ is estimated form the obtained fitting functions.
All the liquids with a secondary process have been treated in this way. Note that analyzing
the data in this manner we are implicitly assuming TTS. 
A final remark is that if we compute $\chi_4$ from the original (normalized) relaxation function, instead that from the fitting functions, we find a relative difference beteween the heights of the maxima only of a few percents.

\end{document}